\documentclass[12]{iopart}
\usepackage{epsfig}

\begin{document}
\title{Quantum nonlocality of Heisenberg XX model with Site-dependent Coupling Strength}
\author{Chunfeng Wu\dag, Jing-Ling Chen\dag, D. M. Tong\dag, L. C. Kwek\dag\ddag, and C. H. Oh\dag }

\address{\dag\ Department of Physics, National University of
Singapore, 2 Science Drive 3, Singapore 117542} 
\address{\ddag\ Nanyang Technological University, National Institute
of Education, 1, Nanyang Walk, Singapore 637616} 
\ead{g0201819@nus.edu.sg}

\begin{abstract}
We show that the generalized Bell inequality is violated in the
extended Heisenberg model when the temperature is below a
threshold value. The threshold temperature values are obtained by
constructing exact solutions of the model using the
temperature-dependent correlation functions. The effect due to the
presence of external magnetic field is also illustrated.
\end{abstract}
 \maketitle

\section {Introduction}
An intriguing aspect of quantum mechanics is the lack of a local
realistic description that could reproduce the necessary
correlations for the experimental outcomes in composite
systems\cite{EPR}. This lack of local realism can be investigated
using the entangled state as discussed in the original seminal
paper by Einstein, Podolsky and Rosen. Nowadays, we recognize the
importance of entanglement as a valuable resource for quantum
information processing and communication. Its usefulness has
since been demonstrated clearly in processes like quantum
teleportation\cite{CHB, TCR}, quantum computation\cite{Deutsch},
and quantum cryptography\cite{Ekert}.

However, concepts such as entanglement and its implications
concerning the non-existence of a local realism in quantum
mechanics have a more fundamental role in quantum mechanics. The
issue of "locality" as well as notion of quantum measurements has
given rise to some of the recent and modern interpretations of
quantum mechanics as well as a better understanding of quantum
phenomena\cite{Peres}. It is also amidst all these theoretical
constructs that Bell proposed an inequality that could rule out
the hidden variable description of quantum mechanics\cite{Bell}.
Since then, several variants of Bell inequalities, some of which
were more amenable for experimental investigations, have been
derived for two-body correlation functions to investigate the
existence of local realism\cite{nqubit}.

Recently there has been much work on the implementation of
quantum processing on solid state devices. In this paper, we study
the thermal states in a system of interaction spins and
investigate its quantum ``nonlocality". An interesting type of
entanglement, thermal entanglement, was studied in the context of
the Heisenberg XXX\cite{XXX,XXX1}, XX\cite{XX}, and XXZ\cite{XXZ}
models. The Heisenberg model has been shown to have a potential
candidate as a model for spin-spin interaction in a solid state
quantum computer\cite{QC}. Being the large Coulomb repulsion
limit of the Hubbard model, it has been partially realized in
quantum dots\cite{QC}, nuclear spins\cite{NS}, and optical
lattices\cite{OL}. In a recent work, Imamoglu \emph{et
al}\cite{QD} have realized quantum information processing using
quantum dot spins and cavity QED, and obtained an effective
interaction Hamiltonian based on the XY spin chain between two
quantum dots. The effective Hamiltonian was shown to be capable
of constructing the Controlled-Not gate\cite{QD}. The XY
Hamiltonian is given by
\begin{eqnarray}
H=\sum_{n=1}^{N}(J_1S_n^xS_{n+1}^x+J_2S_n^yS_{n+1}^y)
\end{eqnarray} where $S^i=\sigma^i/2 (i=x,y,z)$ and $\sigma^i$ are Pauli
operators. When $J_1=J_2$, the XY model becomes XX model. In the
XY model, the interaction strength between neighboring sites is
usually assumed to be independent of the sites. In most solid
state models however, the inter-site coupling strength is site
dependent. In this paper we consider an extended quantum XX model
in which the interaction strength assumes a particular site
dependent form.

This paper is organized as follows. In Sec.\ref{20}, solutions of
the extended XX model for 4 particles are given. In Sec.\ref{30},
we construct the temperature dependent correlation functions in
terms of thermal equilibrium state and investigate the violation
of Bell inequality for the thermal state. The threshold
temperature is given. We also point out that the eigenstates of
the extended XX model do not realize maximal violation of Bell
inequality. Effect of external magnetic field is discussed in
Sec.\ref{40} and we end with some discussions in the final
section.

\section {Solution of the extended XX model}
\label{20} The extended XX Heisenberg model is described by the
Hamiltonian
\begin{eqnarray}
H&&=2\sum_{n=1}^{N-1}J_{n,n+1}(\sigma_n^x\sigma_{n+1}^x+\sigma_n^y\sigma_{n+1}^y)
\nonumber  \\&&
=\sum_{n=1}^{N-1}J_{n,n+1}(\sigma_n^+\sigma_{n+1}^-+\sigma_n^-\sigma_{n+1}^+)
\end{eqnarray}
where $J_{n,n+1}=\sqrt{n(N-n)}$ is the coupling strength between
lattices $n$ and $n+1$. Obviously, the Hamiltonian $H$ describes a
nearest-neighbor interaction spin chain. Interestingly, such a
Hamiltonian has been shown to be useful for perfect state
transfer in quantum spin networks \cite{artur}. The Hamiltonian
$H$ possesses $2^N$ complete and orthonormal eigenstates.

When spin chains are subjected to environmental disturbance, they
inevitably become thermal equilibrium states. The state of a
system at finite temperature $T$ is given by the Gibb's density
operator $\rho(T)=\exp(-H/kT)/Z$, where $Z={\rm Tr}[\exp(-H/kT)]$
is the partition function, $H$ is the system Hamiltonian and $k$
is the Boltzmann constant, which is set to unity for convenience
in this paper. At high temperature, the thermal state becomes
maximally mixed and do not violate Bell inequalities of  any
kind. It is therefore interesting to consider the critical
temperature at which a Bell inequality will be violated. For a
two-qubit system, we have the original Bell inequality.  For
arbitrary number of qubits, we have the Zukowski-Brukner
inequality\cite{nqubit}.

Unfortunately it is not possible to test Zukowski-Brukner
inequality for three qubits in this case since the correlation
functions defined below are zero. Therefore, in this paper, we
first focus on the next non-trivial case of a 4-qubit system and
test the violation of local realistic description using the
Zukowski-Brukner inequality. The extension to arbitrary number of
sites, albeit complicating, can also be done in the same manner.
The Hamiltonian has sixteen eigenvalues
\begin{eqnarray}
&&E_0=E_7=E_8=E_{15}=0,   \nonumber  \\
&& E_3=E_{13}=-1, \;\; E_4=E_{14}=1, \nonumber\\
&&E_6=-2, \;\; E_9=2,     \nonumber  \\
&&E_1=E_{11}=-3, \;\; E_2=E_{12}=3, \nonumber\\
&&E_5=-4,\;\; E_{10}=4.
\end{eqnarray}
The corresponding eigenstates $\{ |\phi_0\rangle, |\phi_1\rangle,
\cdots |\phi_{15}\rangle \}$ can be computed easily and can be found
in appendix \ref{append1}. The above eigenvalues and eigenstates
completely determine the thermal states. The density operator
$\rho(T)$ at the temperature $T$ can be written as
\begin{eqnarray}
\rho(T)=\frac{1}{Z}\sum_{\mu=0}^{15}e^{-\beta
E_{\mu}}|\phi_{\mu}\rangle\langle\phi_{\mu}|
\end{eqnarray}
where $\beta=1/T$ and the partition function
\begin{eqnarray}
Z&&={\rm Tr}(e^{-\beta H})=\sum_{\mu=0}^{15}e^{-\beta E_{\mu}} \nonumber  \\
&&=4+4\cosh(3\beta)+4\cosh\beta+2\cosh(4\beta)+2\cosh(2\beta)
\end{eqnarray}

\section{Violation of 4-qubit Bell inequality and the threshold temperature}
\label{30} To test quantum nonlocality for the state $\rho(T)$,
correlation function $Q_{ijkl}$ should be computed. From the
definition of $Q_{ijkl}$\cite{nqubit}, we have
\begin{eqnarray}
Q_{ijkl}&&={\rm Tr}[\rho(\hat{n}_i\cdot\vec{\sigma})\otimes(\hat{n}_j\cdot\vec{\sigma})\otimes(\hat{n}_k\cdot\vec{\sigma})\otimes(\hat{n}_l\cdot\vec{\sigma})]\nonumber \\
&&=\frac{1}{Z}\sum_{\mu=0}^{15}e^{-\beta E_{\mu}}{\rm Tr}[|\phi_{\mu}\rangle\langle\phi_{\mu}|(\hat{n}_i\cdot\vec{\sigma})\otimes(\hat{n}_j\cdot\vec{\sigma})\otimes(\hat{n}_k\cdot\vec{\sigma})\otimes(\hat{n}_l\cdot\vec{\sigma})] \nonumber \\
&&=\frac{1}{Z}\sum_{\mu=0}^{15}e^{-\beta E_{\mu}}Q^{\mu}_{ijkl}
\label{eq7}
\end{eqnarray}
where $\hat{n}_\alpha=(\sin\theta_{\alpha},0,\cos\theta_{\alpha}),
\ \alpha=i,j,k,l$. $Q^{\mu}_{ijkl}$ is the correlation function
for the eigenstate $|\phi_{\mu}\rangle$,
\begin{eqnarray}
Q^{\mu}_{ijkl}={\rm
Tr}[|\phi_{\mu}\rangle\langle\phi_{\mu}|(\hat{n}_i\cdot\vec{\sigma})\otimes(\hat{n}_j\cdot\vec{\sigma})\otimes(\hat{n}_k\cdot\vec{\sigma})\otimes(\hat{n}_l\cdot\vec{\sigma})]
\end{eqnarray}
For instance, the quantum correlation for the ground state
$|\phi_5\rangle$ is given by
\begin{eqnarray}
Q^5_{ijkl}=&&\cos\theta_i\cos\theta_j\cos\theta_k\cos\theta_l+\frac{\sqrt{3}}{2}\cos\theta_k\cos\theta_l\sin\theta_i\sin\theta_j \nonumber \\
&&-\frac{\sqrt{3}}{4}\cos\theta_j\cos\theta_l\sin\theta_i\sin\theta_k+\frac{1}{2}\cos\theta_i\cos\theta_l\sin\theta_j\sin\theta_k \nonumber \\
&&+\frac{1}{2}\cos\theta_j\cos\theta_k\sin\theta_i\sin\theta_l-\frac{\sqrt{3}}{4}\cos\theta_i\cos\theta_k\sin\theta_j\sin\theta_l \nonumber \\
&&+\frac{\sqrt{3}}{2}\cos\theta_i\cos\theta_j\sin\theta_k\sin\theta_l
+\sin\theta_i\sin\theta_j\sin\theta_k\sin\theta_l.
\end{eqnarray}
Other quantum correlation functions can also be calculated in a
similar way. The correlation function for the thermal state
$\rho(T)$ are computed using Eq.(\ref{eq7}). Based on the
calculated values of $Q_{ijkl}$, we construct Bell quantity ${\cal
B}$
\begin{eqnarray}
{\cal B}=&&Q_{1111}-Q_{1112}-Q_{1121}-Q_{1122}-Q_{1211}-Q_{1212} \nonumber  \\
&&-Q_{1221}+Q_{1222}-Q_{2111}-Q_{2112}-Q_{2121}+Q_{2122}  \nonumber  \\
&&-Q_{2211}+Q_{2212}+Q_{2221}+Q_{2222}
\end{eqnarray}

\begin{figure}
\begin{center}
\epsfig{figure=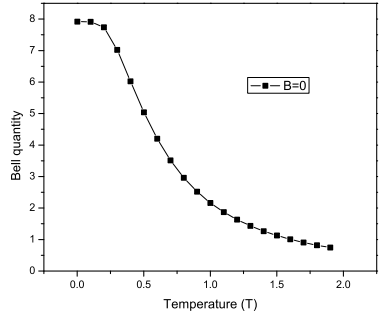,width=0.45\textwidth}
\end{center}
\caption{For a local realistic description of quantum mechanics,
the Bell quantity ${\cal B}$ must necessarily be less than 4.
However, the Bell quantity as a function of temperature $T$ shows
that there is a significant violation of Bell inequality at $T <
0.626$.} \label{fig1}
\end{figure}

For a local realistic description, we require $-4\leq{\cal
B}\leq4$. In Figure \ref{fig1}, we have numerically computed the
Bell quantity as a function of temperature. The results show that
violation of the Bell inequality occurs at $T\leq T_0=0.626$. We
call this critical value $T_0$ the threshold temperature. The
maximum value of ${\cal B}$ for the state $\rho(T)$ approaches
$7.917$ at temperature close to zero.

We have also evaluated the Bell quantity ${\cal
B}(|\phi_{\mu}\rangle)$ in terms of correlation functions of each
pure state $|\phi_{\mu}\rangle$. The maximum value of ${\cal
B}(|\phi_{\mu}\rangle)$ are
\begin{eqnarray}
{\cal B}_{max}(|\phi_{\mu}\rangle)= &&4 \ \ \ \ \ \ \ \ \ \  {\rm for} \ |\phi_{0,15}\rangle  \nonumber \\
                            &&6.112 \ \ \ \ \   {\rm for} \ |\phi_{1,2,3,4,11,12,13,14}\rangle  \nonumber  \\
                            &&7.917 \ \ \ \ \ {\rm for} \ |\phi_{5,10}\rangle  \nonumber  \\
                            &&5.657 \ \ \ \ \ {\rm for} \ |\phi_{6,9}\rangle  \nonumber  \\
                            &&4.866 \ \ \ \ \ {\rm for} \ |\phi_{7}\rangle  \nonumber  \\
                            &&4.060 \ \ \ \ \ {\rm for} \ |\phi_{8}\rangle  \nonumber  \\
\end{eqnarray}

We can explain qualitatively why the maximum value of ${\cal B}$
for the thermal state should be $7.917$ by noting that the
thermal state $\rho(T)$ is the linear combination of
$|\phi_{\mu}\rangle\langle\phi_{\mu}|$ weighted with the factors
$e^{-\beta E_{\mu}}$. For eigenvalue $E_5=-4$, ${\cal
B}_{max}(|\phi_5\rangle)=7.917$, the power is $e^{4\beta}$ and
when $\beta$ is large enough, the Bell quantity ${\cal B}$ is
totally determined by the contribution of state $|\phi_5\rangle$.
Another thing worth noting is that the eigenstates of extended XX
model do not lead to highest value of ${\cal B}_{max}$. We check
the maximum value of the Bell quantities consist of correlation
functions for the following three general states
%\begin{widetext}
\begin{eqnarray}
|\phi'\rangle=&&\cos\alpha_1|1000\rangle+\sin\alpha_1\cos\alpha_2|0100\rangle  \nonumber  \\
&&+\sin\alpha_1\sin\alpha_2\cos\alpha_3
|0010\rangle+\sin\alpha_1\sin\alpha_2\sin\alpha_3|0001\rangle
\end{eqnarray}
\begin{eqnarray}
|\phi''\rangle=&&\cos\alpha_1|1110\rangle+\sin\alpha_1\cos\alpha_2|1101\rangle  \nonumber  \\
&&+\sin\alpha_1\sin\alpha_2\cos\alpha_3
|1011\rangle+\sin\alpha_1\sin\alpha_2\sin\alpha_3|0111\rangle
\end{eqnarray}
\begin{eqnarray}
|\phi'''\rangle=&&\cos\alpha_1|1100\rangle+\sin\alpha_1\cos\alpha_2|1010\rangle+\sin\alpha_1\sin\alpha_2\cos\alpha_3|1001\rangle \nonumber  \\
&&+\sin\alpha_1\sin\alpha_2\sin\alpha_3\cos\alpha_4|0110\rangle+\sin\alpha_1\sin\alpha_2\sin\alpha_3\sin\alpha_4\cos\alpha_5|0101\rangle \nonumber  \\
&&+\sin\alpha_1\sin\alpha_2\sin\alpha_3\sin\alpha_4\sin\alpha_5|0011\rangle
\end{eqnarray}
%\end{widetext}
and find that
\begin{eqnarray}
&&{\cal B}_{max}(|\phi_0'\rangle)=6.217  \nonumber  \\
&&{\cal B}_{max}(|\phi_0''\rangle)=6.217  \nonumber  \\
&&{\cal B}_{max}(|\phi_0'''\rangle)=8.485   \nonumber  \\
\end{eqnarray}
for
$|\phi'_0\rangle=1/2(|1000\rangle+|0100\rangle+|0010\rangle+|0001\rangle)$,
$|\phi''_0\rangle=1/2(|1110\rangle+|1101\rangle+|1011\rangle+|0111\rangle)$
and
$|\phi'''_0\rangle=1/\sqrt{6}(|1100\rangle+|1010\rangle+|1001\rangle+|0110\rangle+|0101\rangle+|0011\rangle)$
respectively. It is easy to see that the degree of violation of
Bell inequality for state $|\phi'_0\rangle$ is higher than that
for the eigenstates $|\phi_{\mu}\rangle, (\mu=1,2,3,4)$ listed in
Eq. (\ref{eq4}). The same results also happen for the eigenstates
$|\phi_{\mu}\rangle, (\mu=11,12,13,14)$ and $|\phi_{\mu}\rangle,
(\mu=5,6,7,8,9,10)$ respectively. We see that among all possible
${\cal B}_{max}$, the state $|\phi'''_0\rangle$ yields the
largest violation.

\section{The effect of external magnetic field}
\begin{figure}
\begin{center}
\epsfig{figure=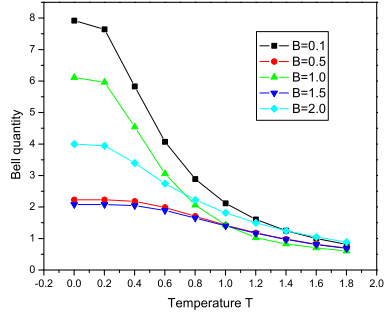,width=0.45\textwidth}
\end{center}
\caption{Bell quantity for the cases with magnetic field $B=0.1,
0.5, 1.0, 1.5$, and $2$.} \label{fig2}
\end{figure}

In this section, we would like to study the effect of magnetic
field on the nonlocality property of thermal state in a general
way, for which the Hamiltonian becomes
\begin{eqnarray}
H'=2
\sum_{n=1}^{N-1}J_{n,n+1}(\sigma_n^+\sigma_{n+1}^-+\sigma_n^-\sigma_{n+1}^+)+B\sum_{n=1}^{N}\sigma_z
\end{eqnarray}
where $B$ is the strength of the magnetic field. It is easy to
verify that the eigenstates of $H'$ are identical with the ones
listed in expression(\ref{eq4}) of $H$, but with different
eigenvalues.
\begin{eqnarray}
&&E'_0=4B, \ \ \ \ E'_1=-3+2B, \ \ E'_2=3+2B, \ \  E'_3=-1+2B,  \nonumber  \\ &&E'_4=1+2B, \ \ E'_5=-4, \ \ \ \ E'_6=-2, \ \ \ \ E'_7=0,  \nonumber  \\
&& E'_8=0, \ \ \ \ E'_9=2, \ \ \ \ E'_{10}=4, \ \ \ \ E'_{11}=-3-2B,    \nonumber\\
&&E'_{12}=3-2B, \ \ E'_{13}=-1-2B, \ \ E'_{14}=1-2B, \ \
E'_{15}=-4B.
\end{eqnarray}
and hence, a new correlation function and Bell quantity ${\cal
B'}$ are given
\begin{eqnarray}
Q'_{ijkl}=\frac{1}{Z'}\sum_{\mu=0}^{15}e^{-\beta
E'_{\mu}}Q^{\mu}_{ijkl}
\end{eqnarray}
\begin{eqnarray}
{\cal B'}=&&Q'_{1111}-Q'_{1112}-Q'_{1121}-Q'_{1122}-Q'_{1211}-Q'_{1212} \nonumber  \\
&&-Q'_{1221}+Q'_{1222}-Q'_{2111}-Q'_{2112}-Q'_{2121}+Q'_{2122}  \nonumber  \\
&&-Q'_{2211}+Q'_{2212}+Q'_{2221}+Q'_{2222}
\end{eqnarray}
where $Z'={\rm Tr}(e^{-\beta H'})$. Now the violation of Bell
inequality depends not only on the temperature, but also on
external magnetic field. Our numerical calculations are shown in
Fig \ref{fig2}.

There are five curves corresponding to $B=0.1, 0.5, 1.0, 1.5$, and
$2$ respectively. When $B=0.1$, the Bell quantity shows a similar
variation of the violation of Bell inequality as a function of
$T$ in the absence of magnetic field. With the increasing value
of external magnetic field, the maximum value of the Bell quantity
approaches the value 2 for which the $B$ field is about $1.5$.
The variation of the Bell quantity as a function of magnetic
field can be explained qualitatively as follows. The $\rho'(T)$ is
a different combination of $|\phi_{\mu}\rangle\langle\phi_{\mu}|$
compared with $\rho(T)$. The largest contribution of all the
states $|\phi_{\mu}\rangle$ is determined by the value of $B$.
When $B<0.5$, it is the eigenstate, $|\phi_5\rangle$, which
ultimately determines the maximal value of the Bell quantity
(${\cal B}_{max}=7.917$) since $e^{-\beta E'_5}=e^{4\beta}$ is the
largest power among all the factors. When $0.5<B<1.5$,
$|\phi_{11}\rangle$ takes the place of $|\phi_5\rangle$ with
power $e^{(3+2B)\beta}$ and ${\cal B}_{max}=6.112$ at $B=1.0$,
for example. When $B>2$, $e^{-\beta E'_{15}}=e^{4B\beta}$ is the
one with largest contribution and ${\cal B}_{max}=4$. But there
are two singular values of $B=0.5$ and $1.5$. In these two cases,
${\cal B}_{max}<4$. The reason for this is that the largest
factors of $e^{-\beta E'_{\mu}}$ are $e^{-\beta E'_5}=e^{-\beta
E'_{11}}=e^{4\beta}$ for $B=0.5$, $e^{-\beta E'_{15}}=e^{-\beta
E'_{11}}=e^{6\beta}$ for $B=1.5$, respectively. Thus the Bell
quantity is determined principally using a combinations of these
two elements of $Q^{\mu}_{ijkl}$, namely,
$e^{4\beta}(Q^5_{ijkl}+Q^{11}_{ijkl})$ and
$e^{6\beta}(Q^{15}_{ijkl}+Q^{11}_{ijkl})$. Note that the maximum
values of the Bell quantity for the latter two correlation
functions are $2.228$ and $2.081$ respectively.

\begin{table}
\begin{tabular}{|c|c|c|c|c|c|c|c|c|c|c|c|c|c|c|c|c|}
     \hline
$B$   &$0$   &$0.1$  &$0.2$  &$0.3$  &$0.4$  &$0.5$  &$0.6$  &$0.7$ \\
     \hline
$T_0$   &$0.626$  &$0.611$  &$0.556$  &$0.447$  &$0.248$  &$None$ &$0.122$  &$0.243$ \\
 \hline
$B$  &$0.8$  &$0.9$  &$1.0$  &$1.1$  &$1.2$  &$1.3$  &$1.4$  &$1.5 \  and  \ above$ \\
\hline
$T_0$  &$0.351$ &$0.427$ &$0.467$  &$0.472$  &$0.436$  &$0.343$  &$0.18$  &$None$\\
\hline
\end{tabular}
\caption{Threshold temperatures for different strengths
of the external magnetic field. When $B=0.5$ and $B=1.5\ and\
above$, the values of Bell quantity are no greater than $4$ at
all times. Therefore, no threshold temperatures exist for these
cases.} \label{tab}

\end{table}

The critical temperatures under different magnetic fields have
been found (Table \ref{tab}). The variation of $T_0$ with increasing
strengths of $B$ is more complicated. This complication arises
mainly because the eigenstates contributing to the optimization
of critical temperatures are different from those needed for the
optimization of magnetic fields. In the latter case, ${\cal
B}_{max}$ is totally determined by the contribution of state with
the largest weight or factor for sufficiently large $\beta$. In
the former case, depending on the value of the external magnetic
field, the eigenstates contributing to the optimization changes
and so the optimization is determined using a combination of the
correlation functions from different states. In short, the
variation of $T_0$ with $B$ is different from that of ${\cal
B}_{max}$ with $B$.

\section {Conclusion}
\label{40} In this paper, we consider the extended Heisenberg XX
model , modeling the nearest-neighbor interaction spin chain. For
the 4-qubit extended XX model, it is shown that since the
correlation functions depend on the temperature and the magnetic
field, the violation of Bell inequality for the thermal state
depends critically on these two parameters. The effect of
temperature for a local realistic description of quantum theory is
determined by the threshold value of $T$ below which the thermal
state violates Bell inequality. The effects of temperature are
also studied at different strengths of magnetic field. For a
fixed temperature, we can find the optimal value of the external
magnetic field that for the violation of Bell inequalities. Our
results imply that quantum ``nonlocality" could be effectively
controlled by magnetic field and temperature. We restrict
ourselves to the 4-qubit case. However, we could also have
discussed the violation of Bell inequality for thermal state for
2-qubit and 3-qubit cases. For 2-qubit extended XX model, the
Bell quantity approaches $2\sqrt2$ which is the maximal violation
of 2-qubit Bell inequality and the corresponding threshold value
of temperature is $T_0=0.667$ when $B=0$. However, for 3-qubit
case, the correlation function defined by this method is always
equal to 0. The violation of Bell inequality for arbitrary number
of qubit can also be done in the same manner.

This work is supported by NUS academic research grant WBS: R-144-000-089-112. J.L.C acknowledges financial support from Singapore Millennium Foundation and (in part) by NSF of China (No. 10201015).

\appendix
\section{Eigenstates of the 4-qubit Hamiltonian}\label{append1}

Corresponding to the sixteen eigenvalues of the Hamiltonian
\begin{eqnarray}
&&E_0=E_7=E_8=E_{15}=0,   \nonumber  \\
&& E_3=E_{13}=-1, \;\; E_4=E_{14}=1, \nonumber\\
&&E_6=-2, \;\; E_9=2,     \nonumber  \\
&&E_1=E_{11}=-3, \;\; E_2=E_{12}=3, \nonumber\\
&&E_5=-4,\;\; E_{10}=4,
\end{eqnarray}
the orthogonal eigenstates are
\begin{eqnarray}
&&|\phi_0\rangle=|0000\rangle \nonumber \\
&&|\phi_1\rangle=\frac{1}{2\sqrt{2}}(-|1000\rangle+\sqrt{3}|0100\rangle-\sqrt{3}|0010\rangle+|0001\rangle)
\nonumber \\
&&|\phi_2\rangle=\frac{1}{2\sqrt{2}}(|1000\rangle+\sqrt{3}|0100\rangle+\sqrt{3}|0010\rangle+|0001\rangle)
\nonumber \\
&&|\phi_3\rangle=\frac{\sqrt{3}}{2\sqrt{2}}(|1000\rangle-\frac{1}{\sqrt{3}}|0100\rangle-\frac{1}{\sqrt{3}}|0010\rangle+|0001\rangle)\nonumber
\\
&&|\phi_4\rangle=\frac{\sqrt{3}}{2\sqrt{2}}(-|1000\rangle-\frac{1}{\sqrt{3}}|0100\rangle+\frac{1}{\sqrt{3}}|0010\rangle+|0001\rangle)\nonumber
\\
&&|\phi_5\rangle=\frac{1}{4}(|1100\rangle-2|1010\rangle+\sqrt{3}|1001\rangle+\sqrt{3}|0110\rangle-2|0101\rangle+|0011\rangle)\nonumber
\\
&&|\phi_6\rangle=\frac{1}{2}(-|1100\rangle+|1010\rangle-|0101\rangle+|0011\rangle)\nonumber
\\
&&|\phi_7\rangle=\frac{\sqrt{3}}{\sqrt{10}}(|1100\rangle-\frac{2}{\sqrt{3}}|1001\rangle+|0011\rangle)\nonumber
\\
&&|\phi_8\rangle=\frac{5}{2\sqrt{10}}(-\frac{\sqrt{3}}{5}|1100\rangle-\frac{3}{5}|1001\rangle+|0110\rangle-\frac{\sqrt{3}}{5}|0011\rangle)\nonumber
\\ &&|\phi_9\rangle=\frac{1}{2}(-|1100\rangle-|1010\rangle
+|0101\rangle+|0011\rangle)\nonumber \\
&&|\phi_{10}\rangle=\frac{1}{4}(|1100\rangle+2|1010\rangle
+\sqrt{3}|1001\rangle+\sqrt{3}|0110\rangle+2|0101\rangle+|0011\rangle)\nonumber\\
&&|\phi_{11}\rangle=\frac{1}{2\sqrt{2}}(-|1110\rangle+\sqrt{3}|1101\rangle-\sqrt{3}|1011\rangle+|0111\rangle)\nonumber  \\
&&|\phi_{12}\rangle=\frac{1}{2\sqrt{2}}(|1110\rangle+\sqrt{3}|1101\rangle+\sqrt{3}|1011\rangle+|0111\rangle)\nonumber  \\
&&|\phi_{13}\rangle=\frac{\sqrt{3}}{2\sqrt{2}}(|1110\rangle-\frac{1}{\sqrt{3}}|1101\rangle-\frac{1}{\sqrt{3}}|1011\rangle+|0111\rangle)\nonumber  \\
&&|\phi_{14}\rangle=\frac{\sqrt{3}}{2\sqrt{2}}(-|1110\rangle-\frac{1}{\sqrt{3}}|1101\rangle+\frac{1}{\sqrt{3}}|1011\rangle+|0111\rangle)\nonumber  \\
&&|\phi_{15}\rangle=|1111\rangle           \nonumber  \\
\label{eq4}
\end{eqnarray}

\section{quantum correlation functions for each pure
states}\label{append2}

The calculation of the quantum correlation functions is
straightforward. In this appendix, we list all the correlation
functions for each eigenstate of the 4-qubit Hamiltonian for easy
reference.
%\begin{widetext}
\begin{table}
\begin{tabular}{| c |l |}
\hline correlation function & explicit expression \\
\hline
 $ Q^0_{ijkl}=Q^{15}_{ijkl}$ & $
\cos\theta_i\cos\theta_j\cos\theta_k\cos\theta_l $\\
\hline  $ Q^1_{ijkl}=Q^{11}_{ijkl}$&
$-\cos\theta_i\cos\theta_j\cos\theta_k\cos\theta_l -
\frac{\sqrt{3}}{4}\cos\theta_k\cos\theta_l\sin\theta_i\sin\theta_j$
\\
& $
+\frac{\sqrt{3}}{4}\cos\theta_j\cos\theta_l\sin\theta_i\sin\theta_k-\frac{3}{4}\cos\theta_i\cos\theta_l\sin\theta_j\sin\theta_k$
\\
& $
-\frac{1}{4}\cos\theta_j\cos\theta_k\sin\theta_i\sin\theta_l+\frac{\sqrt{3}}{4}\cos\theta_i\cos\theta_k\sin\theta_j\sin\theta_l
$ \\
& $
-\frac{\sqrt{3}}{4}\cos\theta_i\cos\theta_j\sin\theta_k\sin\theta_l
$ \\
\hline  $Q^2_{ijkl}=Q^{12}_{ijkl}$ &
$-cos\theta_i\cos\theta_j\cos\theta_k\cos\theta_l+\frac{\sqrt{3}}{4}\cos\theta_k\cos\theta_l\sin\theta_i\sin\theta_j
$ \\
&
$+\frac{\sqrt{3}}{4}\cos\theta_j\cos\theta_l\sin\theta_i\sin\theta_k+\frac{3}{4}\cos\theta_i\cos\theta_l\sin\theta_j\sin\theta_k
$ \\
& $
+\frac{1}{4}\cos\theta_j\cos\theta_k\sin\theta_i\sin\theta_l+\frac{\sqrt{3}}{4}\cos\theta_i\cos\theta_k\sin\theta_j\sin\theta_l
$ \\
&$
+\frac{\sqrt{3}}{4}\cos\theta_i\cos\theta_j\sin\theta_k\sin\theta_l
$ \\
\hline  $ Q^3_{ijkl}=Q^{13}_{ijkl}$ &
$-\cos\theta_i\cos\theta_j\cos\theta_k\cos\theta_l-\frac{\sqrt{3}}{4}\cos\theta_k\cos\theta_l\sin\theta_i\sin\theta_j
$ \\
&$
-\frac{\sqrt{3}}{4}\cos\theta_j\cos\theta_l\sin\theta_i\sin\theta_k+\frac{1}{4}\cos\theta_i\cos\theta_l\sin\theta_j\sin\theta_k
$ \\
& $
+\frac{3}{4}\cos\theta_j\cos\theta_k\sin\theta_i\sin\theta_l-\frac{\sqrt{3}}{4}\cos\theta_i\cos\theta_k\sin\theta_j\sin\theta_l
$ \\
& $
-\frac{\sqrt{3}}{4}\cos\theta_i\cos\theta_j\sin\theta_k\sin\theta_l
$\\
\hline  $ Q^4_{ijkl}=Q^{14}_{ijkl}$ & $
-\cos\theta_i\cos\theta_j\cos\theta_k\cos\theta_l+\frac{\sqrt{3}}{4}\cos\theta_k\cos\theta_l\sin\theta_i\sin\theta_j
$ \\
&
$-\frac{\sqrt{3}}{4}\cos\theta_j\cos\theta_l\sin\theta_i\sin\theta_k-\frac{1}{4}\cos\theta_i\cos\theta_l\sin\theta_j\sin\theta_k
$ \\ & $
-\frac{3}{4}\cos\theta_j\cos\theta_k\sin\theta_i\sin\theta_l-\frac{\sqrt{3}}{4}\cos\theta_i\cos\theta_k\sin\theta_j\sin\theta_l
$ \\
& $
+\frac{\sqrt{3}}{4}\cos\theta_i\cos\theta_j\sin\theta_k\sin\theta_l
$ \\
\hline  $ Q^6_{ijkl} $ &
$\cos\theta_i\cos\theta_j\cos\theta_k\cos\theta_l+\cos\theta_i\cos\theta_l\sin\theta_j\sin\theta_k
$ \\
&
$-\cos\theta_j\cos\theta_k\sin\theta_i\sin\theta_l-\sin\theta_i\sin\theta_j\sin\theta_k\sin\theta_l
$ \\
\hline  $ Q^7_{ijkl} $ & $
\cos\theta_i\cos\theta_j\cos\theta_k\cos\theta_l+\frac{2\sqrt{3}}{5}\cos\theta_j\cos\theta_l\sin\theta_i\sin\theta_k
$ \\
& $
\frac{2\sqrt{3}}{5}\cos\theta_i\cos\theta_k\sin\theta_j\sin\theta_l
+\frac{3}{5}\sin\theta_i\sin\theta_j\sin\theta_k\sin\theta_l $ \\
\hline  $ Q^8_{ijkl} $ &
$\cos\theta_i\cos\theta_j\cos\theta_k\cos\theta_l+\frac{\sqrt{3}}{10}\cos\theta_j\cos\theta_l\sin\theta_i\sin\theta_k
$ \\
& $
\frac{\sqrt{3}}{10}\cos\theta_i\cos\theta_k\sin\theta_j\sin\theta_l
-\frac{3}{5}\sin\theta_i\sin\theta_j\sin\theta_k\sin\theta_l$ \\
\hline  $ Q^9_{ijkl}$ &
$\cos\theta_i\cos\theta_j\cos\theta_k\cos\theta_l-\cos\theta_i\cos\theta_l\sin\theta_j\sin\theta_k
$ \\
&$+\cos\theta_j\cos\theta_k\sin\theta_i\sin\theta_l-\sin\theta_i\sin\theta_j\sin\theta_k\sin\theta_l
$ \\
\hline  $ Q^{10}_{ijkl}$ &
$\cos\theta_i\cos\theta_j\cos\theta_k\cos\theta_l-\frac{\sqrt{3}}{2}\cos\theta_k\cos\theta_l\sin\theta_i\sin\theta_j
$ \\
&
$-\frac{\sqrt{3}}{4}\cos\theta_j\cos\theta_l\sin\theta_i\sin\theta_k-\frac{1}{2}\cos\theta_i\cos\theta_l\sin\theta_j\sin\theta_k
$ \\
&
$-\frac{1}{2}\cos\theta_j\cos\theta_k\sin\theta_i\sin\theta_l-\frac{\sqrt{3}}{4}\cos\theta_i\cos\theta_k\sin\theta_j\sin\theta_l
$ \\
& $
-\frac{\sqrt{3}}{2}\cos\theta_i\cos\theta_j\sin\theta_k\sin\theta_l
+\sin\theta_i\sin\theta_j\sin\theta_k\sin\theta_l $ \\ \hline
\end{tabular}
\end{table}

\end{document}